\documentclass{article}
\usepackage{spconf,amsmath,graphicx}

\usepackage{hyperref}
\usepackage{adjustbox}
\usepackage{amssymb}
\usepackage{amsfonts}
\usepackage{multirow}
\usepackage{enumitem}
\usepackage{algorithmic}
\usepackage{color,soul}
\setlist{nosep, leftmargin=14pt}

\usepackage{mwe} 

\title{PSO-NET: DEVELOPMENT OF AN AUTOMATED PSORIASIS ASSESSMENT SYSTEM USING ATTENTION-BASED INTERPRETABLE DEEP NEURAL NETWORKS}
\name{Sharif A. Kamran, Molly V. Lucas, Brendon Lutnick, Chaitanya Parmar, Basudha Pal, Asha Patel Shah}
\dname{David Apfel, Steven Fakharzadeh, Lloyd Miller, Stephen Yip, Kristopher Standish, Gabriela Oana Cula}
\address{Johnson \& Johnson Innovative Medicine}
\begin{document}
%
\maketitle
\begin{abstract}
Psoriasis is a chronic skin condition that requires long-term
treatment and monitoring. Although, the Psoriasis Area and
Severity Index (PASI) is utilized as a standard measurement
to assess psoriasis severity in clinical trials, it has many
drawbacks such as (1) patient burden for in-person clinic
visits for assessment of psoriasis, (2) time required for
investigator scoring and (3) variability of inter- and intra-rater
scoring. To address these drawbacks, we propose a novel and
interpretable deep learning architecture called PSO-Net,
which maps digital images from different anatomical regions
to derive attention-based scores. Regional scores are further
combined to estimate an absolute PASI score. Moreover, we
devise a novel regression activation map for interpretability
through ranking attention scores. Using this approach, we
achieved inter-class correlation scores of 82.2\% [95\% CI: 77-
87\%] and 87.8\% [95\% CI: 84-91\%] with two different
clinician raters, respectively.	
\end{abstract}
\begin{keywords}
Psoriasis, Deep Learning, Attention, PASI, Convolutional Neural Networks
\end{keywords}
\section{Introduction}
\label{sec:intro}
Psoriasis (PsO) is a chronic inflammatory skin disease that affects approximately 2\%-3\% of the general population worldwide, and its pathogenesis derives from dysregulation of the immune system. With appropriate and timely treatment, individuals can maintain a high quality of life. In recent years, advanced therapies, including biologics and novel oral therapies, have greatly improved care, especially in moderate-to-severe patients. Clinical trial participation and effective measurement of PsO severity in trials is critical to the continued development of such life-changing therapies \cite{berth2006study,zhou2015dermatology,riaz2024negative}. Remote photography (i.e., photos taken at home spanning the full body) coupled with automated image scoring has been identified as one approach for reducing patient and trial burden, as well as limiting inter- and intra-rater variability in clinical scoring. In the existing literature, several automated systems have been proposed for assessing PsO severity \cite{li2020psenet,huang2023artificial,schaap2022image,lucas2024078}. However, each of these systems has limitations, such as, the need for manual removal of any background objects \cite{li2020psenet,huang2023artificial,schaap2022image}, manual masking of images for attention map generation \cite{huang2023artificial}, and/or exclusion of certain anatomical regions typically utilized by clinicians when evaluating PsO disease severity \cite{schaap2022image,xing2024deep}. Some studies also masked objects and removed patient clothing to focus solely on the skin region \cite{li2020psenet,huang2023artificial,schaap2022image}.

To overcome these drawbacks, we propose PSO-Net, a novel deep learning AI algorithm capable of using any set of remotely captured images from the different PASI-assessed anatomical regions (head \& neck, upper extremities, lower extremities and trunk) to generate an attention-based scoring index, that can further be translated into the anatomical score components of the PASI. The model was trained on 28,060 images collected from more than two hundred patients from all six different Fitzpatrick skin tones (Type I; fair skin to Type VI: dark skin) using photos captured by patients at home with a mobile phone application. The model has three key components. The first component is the Encoder Module, which is responsible for extracting dense features from a set of input images. To enhance the model's performance, we use imagenet \cite{deng2009imagenet} pre-trained encoders. These pre-trained encoders were trained on a large dataset of images depicting humans in various scenarios. The second component is the Attention Mechanism, which allows the model to focus on specific areas within the set of input images that are more informative for assessment of PsO. This helps to highlight the specific lesions or image regions that contribute most to the sub-scores used for calculating the overall PASI score. And the third component is the PSO-Net, which utilizes Regression Activation Map Generation to rank the attention scores to generate saliency heat-maps on top of the body images, highlighting areas of importance for estimating the PASI score. Compared to manual calculation of PASI scores, PSO-Net offers consistent and objective measurements in a significantly more time-efficient manner. Patients simply upload a set of images of their PsO lesions and can receive a severity score within seconds. This reduces the need for dermatologists to evaluate and score patients in person. As a result, clinical trials can monitor patient progress without patients having to visit the clinic, which would be especially advantageous for patients residing in geographic areas with limited access to appropriate dermatological care.

\begin{figure*}
	\centering
	\includegraphics[width=0.7\textwidth]{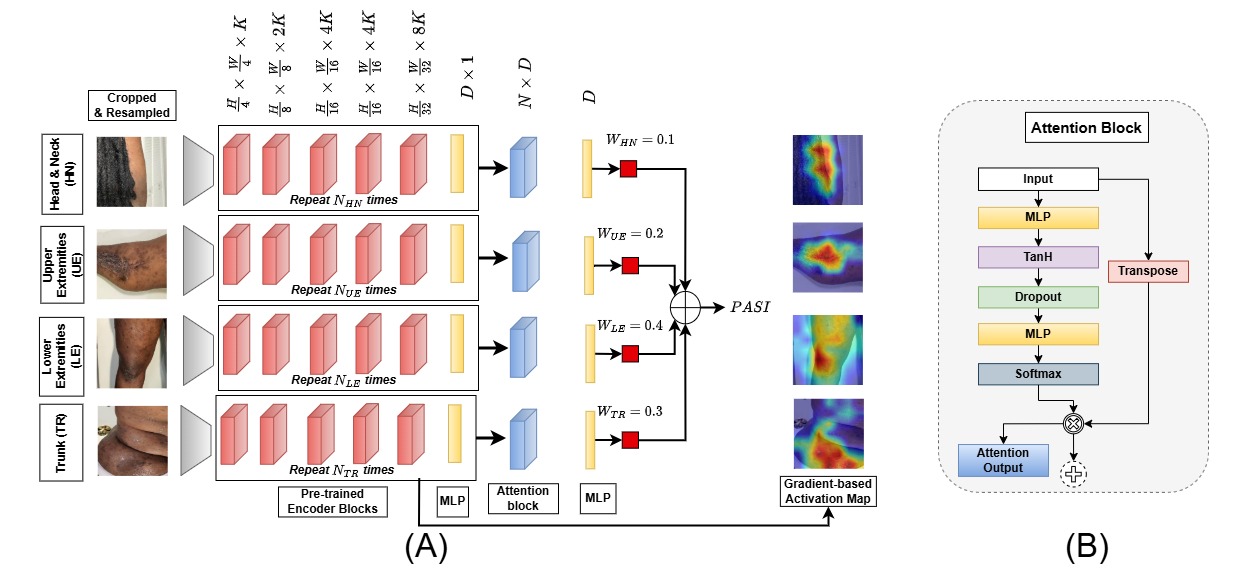}
	\caption{\textbf{(A)} PSO-Net architecture for regional scoring and \textbf{(B)} Attention block used to generate GRAD-RAM}
	\label{fig1}
\end{figure*}
\section{Methodology}
\subsection{PASI Calculation}
The PASI is a standard scoring method for assessing PsO severity and response to treatment and is typically used as a primary endpoint measure in clinical trials [7,8]. The PASI evaluates both qualitative features of plaques (erythema [redness], induration [thickness], and desquamation [scaling]) as well as quantitative aspects of disease activity (extent of surface area involved) across four body regions: head \& neck (HN), upper extremities (UE), lower extremities (LE), trunk (TR). The process of calculating a PASI score for a patient involves the following steps:

\begin{equation}
	PASI_{region} = (A_{ery} + A_{ind} + A_{des}) \times A_{aa}
	\label{eq1}
\end{equation}
\begin{equation}
	PASI_{total} = \sum^{region}_{i} W_{region} \times PASI_{region}
	\label{eq2}
\end{equation}

where region signifies the i-th body region; $W_{region} \in \{0.1,0.2,0.4,0.3\}$ signifies the corresponding weight for that
region. $A_{ery},  A_{ind}$ and $A_{des}$ represent the severity sub-scores for erythema, induration, and desquamation, respectively, and $A_{aa}$ denotes percentage of affected surface area. Clinically, $A_{ery},  A_{ind}$ and $A_{des}$ range from 0 to 4 (ordinal); $A_{aa}$ ranges from 0 to 6 (ordinal); and $PASI_{region}$ and $PASI_{total}$ range from 0 to 72 (continuous), with a higher score indicating greater severity of PsO \cite{george2017automatic}. The manual scoring process for PASI has two limitations. First, the assessor needs to estimate 4 scores ($A_{ery}, A_{ind}, A_{des}$ and $A_{aa}$) for each body region, resulting in a total of 16 variables for a single patient. This estimation heavily relies on the assessor’s level of expertise, and both inter-rater and intra-rater variability can be substantial \cite{langley2004evaluating}. Consequently, PASI measurements can yield inconsistent scoring and assessment of PsO severity \cite{fink2018intra}. Second, the process for measuring each of the 16 variables and calculating the overall PASI score is laborious and time-consuming. Consequently, novel, image-based approaches for measuring each variable and calculating the overall PASI score would reduce the burden on assessors and yield more objective and standardized outcomes.

\subsection{Overall Architecture}
The overall architecture of PSO-Net consists of a pre-trained encoder and an attention block followed by a Multi-Layer Perceptron (MLP) layer for regression prediction, (Fig. \ref{fig1}A). As our dataset consists of human whole-body images captured by patients, we found image-net \cite{deng2009imagenet} pretrained encoders to be useful for transfer-learning. We tested several encoder architectures, including Vision Transformers (ViT), NextViT and ConvNeXt. Vision Transformers (ViT) and NextViT, introduce a paradigm shift by employing self-attention mechanisms to capture long-range dependencies and global context, which are often challenging for traditional convolutional networks. ConvNeXt, a modern reinterpretation of convolutional networks, integrates the strengths of both convolutional operations and transformer-like design principles, enhancing the network's ability to model complex patterns while maintaining computational efficiency. Here, the pre-trained encoder for absolute PASI scoring using the ConvNeXt encoder includes the following steps. First, images for each of the four body regions (HN, UE, LE and TR) are pre-processed and then four separate architectures are used for regional PASI scoring. The pre-trained encoder utilizes multiple convolutions and down sampling blocks. For ConvNeXt, the pre-trained encoder takes Height Width, Channel $(H \times W \times C)$ dimensional RGB images and the feature dimensions after each stage are determined by $\mathbb{R}^{\frac{H}{4}\times \frac{W}{4} \times K}, \mathbb{R}^{\frac{H}{8}\times \frac{W}{8} \times 2K}, \mathbb{R}^{\frac{H}{16}\times \frac{W}{16} \times 4K}$ and  $\mathbb{R}^{\frac{H}{32}\times \frac{W}{32} \times 8K}$. Here, $K$ is the feature dimension. Next, a MLP layer reshapes the feature dimension to $\mathbb{R}^{D \times 1}$. This process is repeated N times depending on how many images from each region are used; we use, $N_{HN} = 12, N_{UE} = 18, N_{LE} =13$ and $N_{TR} = 10$. Next the $N$ features are concatenated to form a feature of size $\mathbb{R}^{N \times D \times 1}$ and $D = 768$, which is then inserted into the attention block shown in Fig. \ref{fig1}(B). The output of the attention block is then connected to the final MLP layer which produces a single value, $D = 1$ for the regional PASI score. Each of the four anatomical architectures produces a continuous value (0.0-72.0), and each is then multiplied by
their respective weight factor (given in Eq. \ref{eq2}) to calculate the absolute PASI score (within a range of 0-72).

\begin{figure}
	\centering
	\includegraphics[width=0.6\columnwidth]{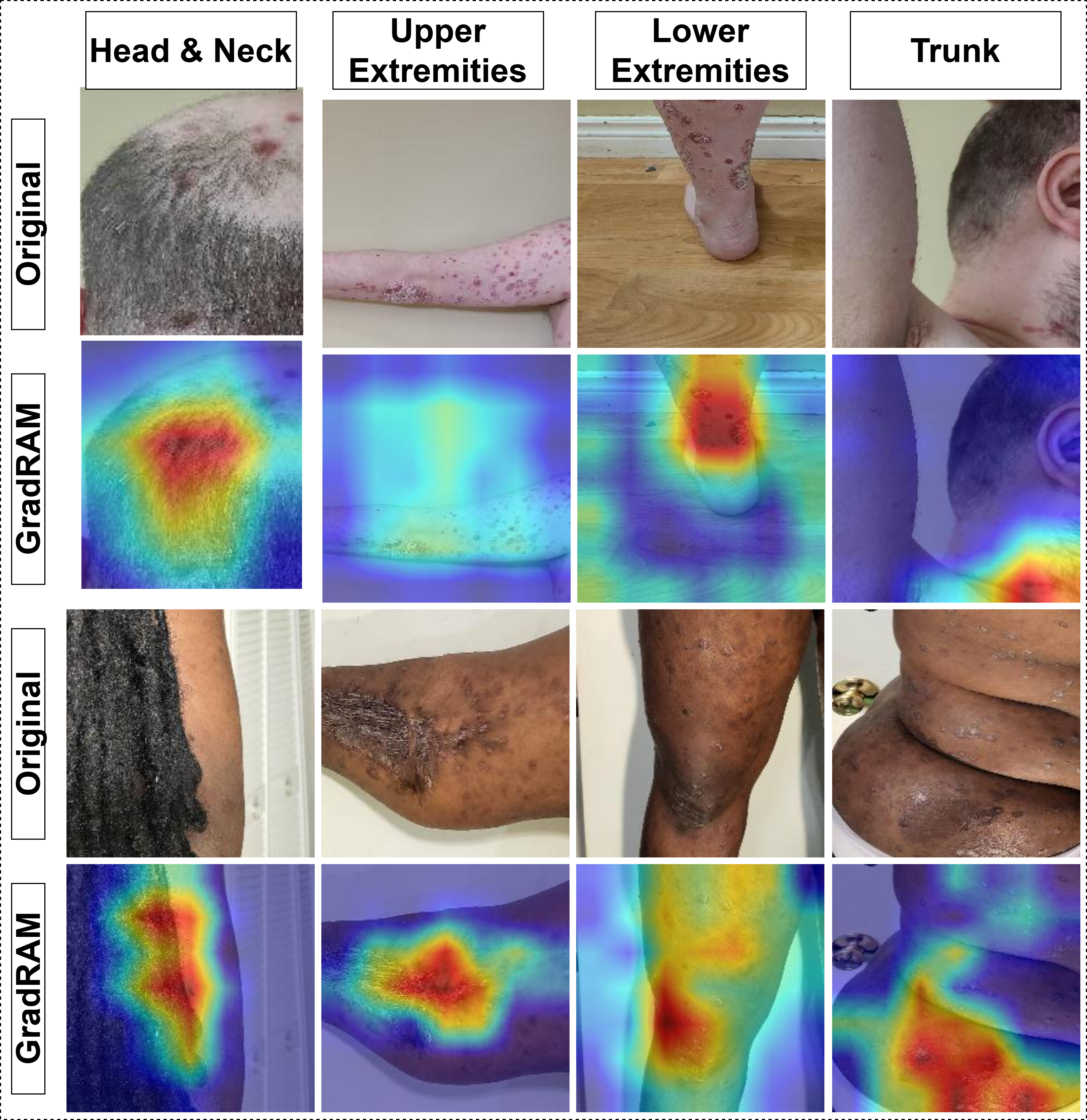}
	\caption{Visualization of Grad-RAM for four regions.}
	\label{fig2}
\end{figure}

\subsection{Attention}
An attention block is designed to capture and emphasize important features. It operates by calculating attention weights, which represent the relative importance of each element of the input with respect to the other elements. For the architecture in Fig. \ref{fig1}(B), we utilize a MLP layer, TanH activation, and then another MLP layer followed by softmax function to determine the attention weight scores for the $N$ images. Attention weights are then multiplied by the output feature vectors and a reduce sum operation is carried out, allowing the model to prioritize and focus on the most informative parts of the input. The input feature vector has a size of $\mathbb{R}^{N\times8K\times1}$ and the output feature size is $\mathbb{R}^{N \times D}$ , where $D=768$ is the dimension of the MLP layer. The attention block is employed for two reasons: 1) to effectively highlight relevant features, enhancing the model's ability to learn and make accurate estimations; and 2) to generate a gradient-based regression activation map (Grad-RAM) between the attention block and MLP layer, as illustrated in Fig \ref{fig1}(A). As there are $N$ different images in each set, separate activation maps are needed for each image. Consequently, a novel, two-step approach for Grad-RAM generation was devised. In the first step, an inference run is performed on the set of images to obtain the regional score and the weighted attention score. A ranking operation is then performed using the weight values to identify images with the highest attention scores. Any specific single image may then be inserted into the regional architecture for back-propagated gradient-based activation map generation, as illustrated in Fig. \ref{fig2}. The resolution of the attention map is $224 \times 224$.

\subsection{Objective Function}
For training purposes, Mean-absolute-error (MAE) Loss was used, as shown in Eq. \ref{eq3}. All regional images are normalized,
to [0, 1] using the image-net normalization process \cite{krizhevsky2012imagenet}. In this context, M is the model that takes x input images and generates an estimation M(x), and y is the ground-truth label rated by the clinician.

\begin{equation}
	\mathcal{L} = \mathbb{E}_{x,y} \parallel M(x) - y \parallel
	\label{eq3}
\end{equation}

\section{Experimentation}
\subsection{Dataset}
Of the 533 patients who provided images, 344 met criteria for inclusion and were advanced through screening. Of these patients, 220 patients were female and 124 were male. In total, 38,824 photos were captured over 844 patient visits (baseline and weeks 2, 4 and 8). Patients completed between 1 and 4 visits. The data collected were divided by patient for purposes of training (70\%; 610/844 timepoints and 247/344 patients), validation (10\%; 64/844 timepoints and 28/344 patients), and use as a test set (20\%; 170/844 timepoints and 69/344 patients). The data set had images from patients across all skin tones as defined by Fitzpatrick Skin Types: I (N=60), II (N=163), III (N=85), IV (N=22), V (N=12) and VI (N=2), which gradually rank skin tones from fair (I) to dark (VI). A
contracted research organization (CRO) was used to collect and rate the image sets for regional and overall PASI scores ($Rater_A$ ). Patients were randomly assigned for assessment by one of seven CRO dermatologists who conducted ratings at baseline and all subsequent visits. We also recruited an eighth dermatologist ($Rater_B$) to re-score the test set with a goal of evaluating inter-rater variability

\subsection{Hyper-parameters}
To train our models we used PyTorch \cite{paszke2019pytorch}. The Adam optimizer \cite{kingma2014adam} was used with a learning rate of $\alpha = 10^{-6}$ , weight decay $= 10^{-4}$, and batch-size of $4$ for $100$ epochs using four NVIDIA A100 GPUs. Weighted sampling was employed to address the patient-based imbalanced PASI distribution (i.e., PASI $> 10$). Two different methods for
experimentation were conducted: 1) with no crops (low resolution); and 2) with 4-crops (high-resolution), where each image was divided into $4$ input images. For the latter, cropping was applied to augment each image-set, increasing the total number of images per-region by $N \times 4$.
\begin{figure}
	\centering
	\includegraphics[width=0.9\columnwidth]{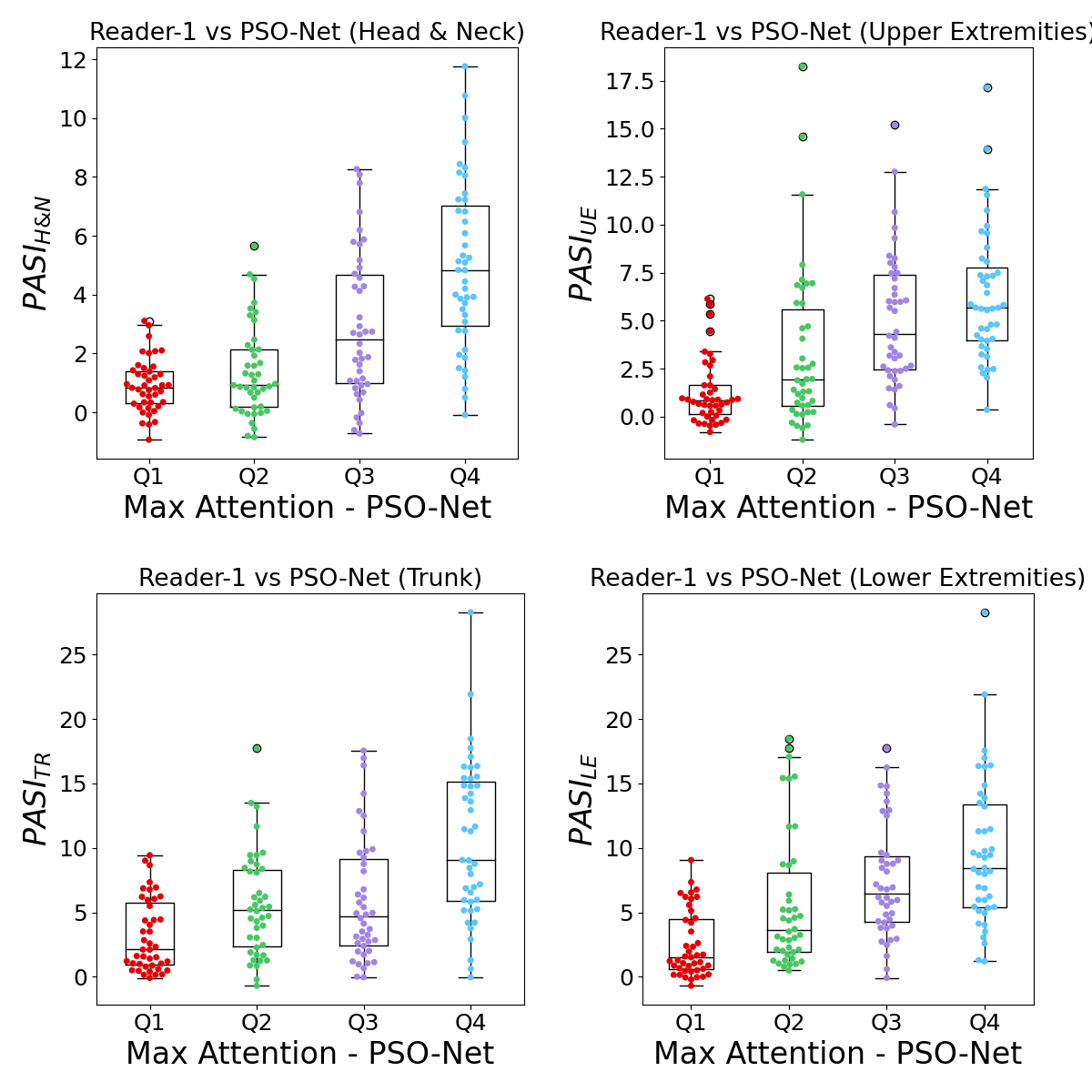}
	\caption{Visualization of max attention generated by the PSONet vs. the PASI regional score.}
	\label{fig3}
\end{figure}

\subsection{Quantitative Evaluation}
For quantitative evaluation of PSO-Net, two metrics were employed to determine Inter-class Correlation Coefficient (ICC) and Mean Absolute Error (MAE). ICC quantifies the inter-rater variability for rater vs model and rater vs rater comparisons, while MAE measures the differences in absolute PASI scores determined by PSO-Net and the assessor dermatologists. As shown in Table \ref{table1}, 4-crop pre-processing in conjunction with the ConvNeXt pre-trained encoder achieved the best ICCs with PSO-Net vs. $Rater_A$ (82.2\%) and PSO-Net vs. $Rater_B$ (87.8\%) and MAEs of 1.68 and 1.59. Conversely, the ViT, ConvNeXt-V2 and NextViT pre-trained encoders underperformed compared to ConvNeXt in terms of both ICC and MAE outcomes. The inter-rater ICC for $Rater_A$ vs. $Rater_B$ was 88.1\%, with a 95\% confidence interval that overlaps with those for the PSO-Net vs. $Rater_A$ and PSO-Net vs. $Rater_B$ ICCs, indicating high correlation.

We also compared our findings against those from other recently published reports \cite{huang2023artificial,xing2024deep,li2020psenet,raj2024objective} incorporating deep learning for PASI estimation. Our model outperformed all prior approaches, except for that described by Raj et. al., with regard to the MAE metric (Table \ref{table2}). Nonetheless, all previously reported models, have limitations, such as lack of regional images, particularly of the head and neck region, reliance on manual ROI for bounding boxes or masks, and inclusion of only one or two demographic groups of patients. PSO-Net overcomes this with attention-based scoring.

\begin{table}
	\caption{Comparison across different architectures and raters
		for absolute PASI scoring.}
	\centering
	\begin{adjustbox}{width=\columnwidth}
		\begin{tabular}{|ccccc|}
			\hline
			\multicolumn{1}{|c|}{Encoder} & \multicolumn{1}{c|}{$ICC_{A}$ [CI95\%]} & \multicolumn{1}{c|}{$MAE_{A} \pm STD$} & \multicolumn{1}{c|}{$ICC_{B}$ [CI95\%]} & $MAE_{B} \pm STD$ \\ \hline\hline
			\multicolumn{5}{|c|}{Low Resolution}                                                                            \\ \hline\hline
			\multicolumn{1}{|c|}{ConvNeXt} & \multicolumn{1}{c|}{$0.781
				[0.71, 0.83]$} & \multicolumn{1}{c|}{$1.90 \pm
				1.93$} & \multicolumn{1}{c|}{$0.853
				[0.81, 0.89]$} & $1.59 \pm
			1.46$ \\ \hline
			\multicolumn{1}{|c|}{NextViT} & \multicolumn{1}{c|}{$0.479
				[0.36, 0.59]$} & \multicolumn{1}{c|}{$2.88
				\pm2.90$} & \multicolumn{1}{c|}{$0.5349
				[0.42, 0.63]$} & $2.61 \pm
			2.74$ \\ \hline
			\multicolumn{1}{|c|}{ViT} & \multicolumn{1}{c|}{$0.783
				[0.72, 0.84]$} & \multicolumn{1}{c|}{$1.89 \pm
				1.89$} & \multicolumn{1}{c|}{$0.835
				[0.78, 0.88]$} & $1.68 \pm
			1.52$ \\ \hline\hline
			\multicolumn{5}{|c|}{High Resolution}                                                                            \\ \hline\hline
			\multicolumn{1}{|c|}{ConvNeXt} & \multicolumn{1}{c|}{$0.822
				[0.77, 0.87]$} & \multicolumn{1}{c|}{$1.68 \pm
				1.81$} & \multicolumn{1}{c|}{$0.878
				[0.84, 0.91]$} & $1.43 \pm
			1.41$ \\ \hline
			\multicolumn{1}{|c|}{NextViT} & \multicolumn{1}{c|}{$0.643
				[0.54, 0.72]$} & \multicolumn{1}{c|}{$2.27 \pm
				2.43$} & \multicolumn{1}{c|}{$0.687
				[0.6, 0.76]$} & $2.21 \pm
			2.06$ \\ \hline
			\multicolumn{1}{|c|}{ViT} & \multicolumn{1}{c|}{$0.789
				[0.73, 0.84]$} & \multicolumn{1}{c|}{$1.86 \pm
				1.85$} & \multicolumn{1}{c|}{$0.843
				[0.79,0.88]$} & $1.65 \pm
			1.47$ \\ \hline
			\multicolumn{1}{|c|}{$Rater_A$} & \multicolumn{1}{c|}{-} & \multicolumn{1}{c|}{-} & \multicolumn{1}{c|}{$0.881
				[0.84, 0.91]$} & $1.41 \pm
			1.57$ \\ \hline
		\end{tabular}
	\end{adjustbox}
	\label{table1}
\end{table}

\subsection{Qualitative Evaluation}
To determine the most informative image lesions and regions for estimating the PASI score for a given patient, we visualized the Grad-RAM and laid it over the corresponding clinical images (Figure \ref{fig2}). Upon inspection, the model identified PsO lesions that more substantially contributed to the overall PASI score. To verify the correlation between attention score and PASI score, we illustrate a swarm plot with mean and quartiles using the test set in Figure \ref{fig3}. The x-axis represents the quartile (Q) of the max attention score per image-set (visits/patients), while the y-axis represents the PASI sub-score for each PASI body region. (HN: $Q1 = 0–0.073, Q2 = 0.073–0.09, Q3 = 0.09–0.122, Q4 > 0.122$; UE: $Q1 = 0–0.065, Q2 = 0.065–0.077, Q3 = 0.077–0.113, Q4 > 0.113$; LE: $Q1 = 0–0.054, Q2 = 0.054–0.068, Q3 = 0.068– 0.09, Q4 > 0.09$; TR: $Q1 = 0–0.081, Q2 = 0.081–0.099, Q3 = 0.099–0.131, Q4 > 0.131$). The model gives higher attention to regions that display high levels of erythema, induration, and scaling. Therefore, max attention scores are correlated with increased PASI and visualized in Figure \ref{fig3}.

\begin{table}
	\caption{Comparisons with previously reported AI-based
		approaches.}
	\centering
	\begin{adjustbox}{width=\columnwidth}
	\begin{tabular}{|c|c|c|c|}
		\hline
		& Model & Limitations & MAE \\ \hline
	Xing et al. \cite{xing2024deep}	& ResNet-34 & No H\&N Poor Stratification &  3.33 \\ \hline
	Huang et al. \cite{huang2023artificial}	& Efficient-Net-B0  & Manual ROI for attention, one demographic & 2.14 \\ \hline
	Li et al. \cite{li2020psenet}	&  PSENets & Single Image Input, manual ROI for detection  & 2.21 \\ \hline
	Raj et al. \cite{raj2024objective}	& MobileNetV2, U-Net & No H\&N, one demographic, manual segmentation &  1.02 \\ \hline
	Ours	& PSO-Net & - &  1.43 \\ \hline
	\end{tabular}
	\end{adjustbox}
	\label{table2}
\end{table}

\section{Conclusion}
In this paper, we propose a novel and interpretable architecture, PSO-Net, which uses the latest advances in AI and deep learning to automatically estimate PASI scores. Moreover, we also devise a novel regression activation map for interpretability by ranking attention scores. Our computer-based model surpasses existing architectures in absolute PASI scoring and exhibits inter-rater variability comparable to what was observed between two separate US board-certified dermatologists. We hope to extend this work and deploy this for real-world clinical trials. 

\section{Acknowledgments}
\label{sec:acknowledgments}
This work was sponsored by Johnson and Johnson.

\section{COMPLIANCE WITH ETHICAL STANDARDS}
This study was performed in accordance with the principles
of the Declaration of Helsinki and was IRB approved.

\bibliographystyle{IEEEbib}
\bibliography{strings,refs}

\end{document}